%% file: cat_tree_doc_retrieval.tex
\documentclass[11pt]{article}
\usepackage{amsmath}
\usepackage{url}
\usepackage{hyperref}

\usepackage{marginnote}
\usepackage{color}
\usepackage{authblk}

\newtheorem{lemma}{Lemma} 
\newtheorem{theorem}{Theorem}

\newtheorem{corollary}{Corollary}
\providecommand{\keywords}[1]{\textbf{\textit{Index terms---}} #1}

\title{Efficient tree-structured categorical retrieval}

\author[1]{Djamal Belazzougui\thanks{Corresponding Author: dbelazzougui@cerist.dz}}
\author[2,3]{Gregory Kucherov\thanks{Gregory.Kucherov@univ-mlv.fr}}
\affil[1]{CAPA, DTISI, Centre de Recherche sur l'Information Scientifique et Technique, Algiers, Algeria.}
\affil[2]{CNRS and LIGM/Univ Gustave Eiffel, Marne-la-Vall\'ee, France.}
\affil[3]{Skolkovo Institute of Science and Technology, Moscow, Russia.}

\begin{document} 
\maketitle

\input{abstract.tex}

\input{Introduction.tex}

\input{preliminaries.tex}

\input{Muthukrishnan_based_solution.tex}

\input{Wavelet_tree_based_solutions.tex}

\input{succinct_compressed.tex}

\input{conclusions.tex}

\bibliography{cat_tree_doc_retrieval.bib}

\end{document}

%% file: abstract.tex
\begin{abstract}
We study a document retrieval problem in the new framework where $D$ text documents are organized in a {\em category tree} with a pre-defined number $h$ of categories. This situation occurs e.g. with taxomonic trees in biology or subject classification systems for scientific literature. Given a string pattern $p$ and a category (level in the category tree), we wish to efficiently retrieve the $t$ \emph{categorical units} containing this pattern and belonging to the category. We propose several efficient solutions for this problem. One of them uses $n(\log\sigma(1+o(1))+\log D+O(h)) + O(\Delta)$ bits of space and $O(|p|+t)$ query time, where $n$ is the total length of the documents, $\sigma$ the size of the alphabet used in the documents and $\Delta$ is the total number of nodes in the category tree. Another solution uses $n(\log\sigma(1+o(1))+O(\log D))+O(\Delta)+O(D\log n)$ bits of space and $O(|p|+t\log D)$ query time. We finally propose other solutions which are more space-efficient at the expense of a slight increase in query time. 
\end{abstract}

%% file: Introduction.tex
\section{Introduction}
Data is often structured using {\em category hierarchies} represented by trees. In many applications, such hierarchies play a crucial guiding role: for example, the International Classification of Diseases (ICD) provides a hierarchical classification of all human disesases and constitues a common reference for diagnostics. In this paper, we are interested in sequence data, such as biological sequences or text documents, that are linked to a given hierarchy. More precisely, in our framework sequences are associated to leaves of a hierarchy, and tree nodes are mapped to several fixed levels, also called {ranks}. 

This situation is common and occurs in several important applications. One is biology where species are classified according to the famous Linnaean taxonomy including eight common {\em taxonomic ranks}: species, genus, family, order, class, phylum, kingdom, domain. Then, given a set of sequences (DNA, RNA or protein) belonging to known species, one can associate them to the corresponding leaves of the taxonomic tree. Such a structure is used, for example, for phylogeny-based metagenomic classification where one considers the tree of known genomic sequences as a reference for classifying sequences of a metagenomic sample, see e.g. \cite{Wood2014}. A classification procedure may involve queries asking for the taxonomic units (i.e. internal nodes of the tree) of a certain rank whose sequences contain a given pattern, or similar type of queries. 

Another example is provided by text documents such as scientific papers. The latter are usually annotated by subjects belonging to a fixed hierarchical nomenclature, such as ACM Computing Classification System (CCS) or Mathematics Subject Classification (MSC). Those subject hierarchies have a predefined number of levels: four levels for CCS and three for MSC. Given a corpus of scientific papers, one could ask about subject categories at a certain level whose documents contain a given pattern. This is a natural information retrieval scenario. 

Here we study this problem from the stringology perspective (see e.g. \cite{KucherovNekrichStarikovskayaSPIRE12,GawrychowskiEtAlSPIRE13}). Assume we are given 
a set of $D$ documents of total length $n$ over an alphabet of size $\sigma$, organized in 
a tree of height $h$. The tree has $D$ leaves, each associated with a distinct document, and the leaves are all at level $h$ of the tree. The total number of nodes in the tree is denoted by $\Delta$. 
The tree specifies a hierarchy of categories: each level of the tree corresponds to a category, and each internal node corresponds to a {\em categorical unit}. 

The basic type of query we study in this paper is the following. 
\begin{quote}
	Given a pattern $p$, and a tree level (rank) $i\in [1..h]$, return all nodes (categorical units) $d_1,\cdots, d_t$ at level $i$ that have at least one leaf (document) in their subtree that contains pattern $p$. 
	\end{quote}
For example, given a large collection of genomic sequences organized in a taxonomic tree (for example, all known animal genomes), one may ask which animal families have a given sequence in the genomes of their members. Or, given a large hierarchy of documents (for example, all Computer Science papers), one may wonder in which subfields of Computer Science (corresponding to a certain level of the hierarchy) the term '{\em suffix tree}' is used. 
This basic type of queries can be further extended in different ways. For example, one may impose an additional requirement of the mimimum number of documents of the categorical unit containing the given pattern. In this first study, we focus on the basic query type. 

In this work, we propose several algorithms for this problem. Our first solution (Section~\ref{muthu-solution}) is based on the approach of Muthukrishnan \cite{muthukrishnan2002efficient} to the document retrieval problem. By combining several algorithmic tools - efficient text index, colored range reporting queries, and level ancestor queries - we obtain a solution with $n(\log\sigma(1+o(1))+\log D+O(h)) + O(\Delta)$ bits of space and $O(|p|+t)$ query time, where $t$ is the output size, i.e. the number of retrieved categorical units. To improve the space bound, in particular to get rid of the $O(nh)$ term which can be as big as $O(nD)$, we then develop a solution based on a wavelet tree built on top of the input category tree (Section~\ref{wavelet-tree-based}). On this way, we first obtain a solution taking $n(\log\sigma+\log D)+O(D\log n)$ bits and $O(|p|+t\cdot h\log D)$ query time. We further improve it using the technique of heavy path decomposition, to obtain a solution in $n(\log\sigma(1+o(1))+\log D)+O(\Delta)$ bits of space and $O(|p|+t\log D)$ query time. In the final part of the paper (Section~\ref{succinct}), we focus on solutions using succinct and compressed data structures, on top of the input data. That is, our main goal here is to replace the $n\log D$ bits by respectively $n\log \sigma$ or by $nH_0+o(n\log\sigma)$ in representing the document array. We obtain memory-time trade-offs showing how this goal can be achieved at the price of a slight increase of query time. 

We summarize our main results in the following table.
\begin{center}
\begin{tabular}{ |c|c|c| } 
 \hline
 algorithm & space (bits) & query time \\ \hline
 based on colored  & $n(\log\sigma(1+o(1))+\log D+O(h))$ & $O(|p|+t)$ \\ 
 range queries (Sect.~\ref{muthu-solution}) & $+ O(\Delta)$ & \\\hline
  based on wavelet & $n(\log\sigma(1+o(1))+O(\log D))$  & $O(|p|+t\log D)$ \\ 
  tree (Sect.~\ref{wavelet-tree-based}) & $+O(\Delta)+O(D\log n)$ & \\\hline
 compact space (Sect.~\ref{succinct}) & $O(n\log\sigma)$& $O(|p|+(t+1)\cdot\log^\epsilon n(1+\frac{h}{\log\sigma}))$\\\hline
 compressed space (Sect.~\ref{succinct}) &$nH_k+o(n\log\sigma)+O(D\log n)$ &$O(|p|+t\cdot h\log n(\log\log n)^2)$\\
 \hline
\end{tabular}
\end{center}

%% file: preliminaries.tex
\section{Preliminaries}
We first briefly present main algorithmic tools used by our algorithms. 

\subsection{Level ancestor queries on trees}
Consider a rooted tree. To each node in the tree we associate its {\em level} so that the level of the root is $1$, and the level of a child node is 1 more than the level of its parent. 
The height of a tree is defined as the maximal level of any node in the tree. We denote by $\ell_\alpha$ the level of a node $\alpha$.

We will use the implementation of level ancestor queries specified by the following lemma. 
\begin{lemma}[\cite{navarro2014fully}]\label{lemma:laq}
There exists a data structure that represents a tree with $n$ nodes within space $2n+o(n)$ and allows answering the following queries in constant time: 
\begin{enumerate}
\item given a level $\ell$ and a node $\alpha$ at level at least $\ell$, return the ancestor node $\beta$ of $\alpha$ at level $\ell$,
\item given an integer $i$, return the node $\alpha$ where $\alpha$ is the leaf number $i$ in left-to-right order. 
\end{enumerate}
\end{lemma}

We denote by $\mathtt{LAQ}(\alpha,i)$ the query which asks for the ancestor at level $i$ of node $\alpha$.
We denote by $\mathtt{leafselect}(i)$ the query which returns the $i$-th leaf of the tree in left to right order.  

\subsection{$\mathtt{rank}$/$\mathtt{select}$ queries and wavelet trees}
\label{bitvecs}
$\mathtt{rank}$ and $\mathtt{select}$ queries on sequences constitute basic building blocks of many succinct data structures  \cite{Jacobson89}. Given a string $S[1..n]$ on an alphabet $\Sigma$, a query $\mathtt{rank}_c(S,i)$, with $c\in\Sigma$ and $i\in[1..n]$, asks for the number of occurrences of $c$ in $S[1..i]$ and 
$\mathtt{select}_c(S,j)$ asks for the unique position $i$ such that $S[i]=c$ and $\mathtt{rank}_c(S,i)=j$. 

Consider first the important case of binary sequences (bitvectors). 
The following result is well-known, see \cite{navarro_compact_2016}.

\begin{lemma}\label{lemma:bitvector}
A bitvector $B[1..n]$ can be represented using $n+o(n)$ bits of space, so that queries $\mathtt{rank}$ and $\mathtt{select}$ are answered in constant time. 
\end{lemma}

In the case of non-binary alphabet, $\mathtt{rank}$/$\mathtt{select}$ queries can be efficiently answered using {\em wavelet trees}. 
The wavelet tree has been formally introduced in~\cite{grossi2003high}, but a similar structure has been used 
earlier \cite{chazelle1988functional}. 
Suppose we are given a sequence $S$ of length $n$ over an alphabet $\Sigma$.

The {\em (binary) wavelet tree} is a binary tree representation of $S$ that is defined recursively as follows. 
Let $\Sigma_0\neq \emptyset$ and $\Sigma_1\neq \emptyset$ form a partition of $\Sigma$ (that is, $\Sigma=\Sigma_0\cup \Sigma_1$ and $\Sigma_0 \cap \Sigma_1=\emptyset$). Then the root of the binary wavelet tree will contain a binary vector $B$, such that $B[i]=0$ iff $S[i]\in \Sigma_0$.
Let the sequence $S_0$ (resp., $S_1$) be formed by keeping  only the elements of $S$ that belong to $\Sigma_0$ (resp., $\Sigma_1$), in the same order. Then, the left (resp., right) child is defined recursively using $S_0$ (resp., $S_1$) and a binary partition of $\Sigma_0$ (resp., $\Sigma_1$). 
The recursion stops whenever we reach a leaf that corresponds to a singleton subset of $\Sigma$. Such nodes will form the leaves of the wavelet tree. We refer the reader to the survey~\cite{navarro2014wavelet} for more details about wavelet trees. 
We will make use of the following lemma: 

\begin{lemma}[\cite{grossi2003high}]\label{lemma:WT}
The wavelet tree over the alphabet $[1..\sigma]$ can be represented using $n(\log\sigma+o(1))+O(\sigma\log n)$ bits of space, supporting 
$\mathtt{rank}$ and $\mathtt{select}$ queries in $O(\log\sigma)$ time. 
\end{lemma}

\mathchardef\mhyphen="2D
\newcommand{\rangedistinct}{\mathtt{range}\mhyphen\mathtt{distinct}}

%

The definition of binary wavelet tree can be readily generalized to the non-binary case. As in the binary case, to any node $\alpha$ labeled by an interval $\Sigma_\alpha$ is (implicitly) associated the sequence $S_\alpha$ which is the subsequence of $S[1..n]$ consisting of all characters belonging to $\Sigma_\alpha$. If a node $\alpha$ of a wavelet tree has $d$ children, then the alphabet interval $\Sigma_\alpha \subseteq[1..\sigma]$ assigned to $\alpha$ is partitioned into $d$ disjoint subintervals instead of two, and $\alpha$ stores a sequence $C_\alpha$ over alphabet $[1..d]$ of length $|S_\alpha|$ such that $C_\alpha[i]=j$ iff $S_\alpha[j]\in \Sigma_{\alpha_j}$.


\subsection{Text indexes}
We assume familiarity with main text indexing structures: suffix trees, suffix arrays and BWT-indexes. Here we only recall some basic facts about them. 

Given a text $T$ over an alphabet $\Sigma=[1..\sigma]$, a suffix tree~\cite{weiner1973linear} is a tree data structure that stores in its leaves the suffixes of $T\$$, where $\$$ is a special character that does not appear in $T$ and is lexicographically smaller than any character of $T$. Each suffix is associated with its starting position in $T\$$. Suffix tree allows answering basic string pattern matching queries: given a pattern $p$, return the set of starting positions of $p$ in $T$. 
%

The suffix array of $T$ is a related but more space-efficient data structure defined as the array $\mathtt{SA}[1..n+1]$ obtained by sorting all the suffixes of $T\$$ in lexicographic order and setting $\mathtt{SA}[i]=j$ if and only if the suffix $T[j..n]\$$ has lexicographic rank $i$ among all suffixes of $T\$$. 

A suffix tree occupies $O(n\log n)$ bits of space and a matching query needs access to the original text $T$ in addition to the suffix tree. The query time is $O(|p|\log\sigma)$. The suffix array~\cite{manber1993suffix} is an alternative to the suffix tree which occupies the same $O(n\log n)$ bits of space, but has lower constant factors in space and supports matching queries in $O(|p|+\log n)$ time. 

The BWT-index (FM-index) is a space-efficient alternative to suffix arrays and suffix trees which uses $O(n\log\sigma)$ bits of space only. It was originally proposed in~\cite{ferragina2005indexing} and has seen many improvements. We will use the following version of BWT-index with alphabet-independent query time. 
\begin{lemma}[\cite{belazzougui2014alphabet}]~\label{lemma:textindex}
Given a text $T$ of length $n$ over alphabet $[1..\sigma]$, we can build a BWT-index which occupies $n\log\sigma(1+o(1))$ bits of space and supports computing the range of suffixes prefixed by a pattern $p$ in time $O(|p|)$. 
\end{lemma}
Note that computing the range of suffixes answers also whether the pattern occurs in the text at all, and if so, reports the number of its occurrences (the size of the lexicographic order interval). 
For this reason, the query presented in the lemma above is usually refered to as a $\mathtt{count}$ query. 
The BWT-index is usually augmented with position information so that it becomes able to report the location of each occurrence of the pattern in addition to the number of occurrences. This can be achieved using fo the example the compressed suffix array representation:
\begin{lemma}[\cite{grossi2005compressed}]~\label{lemma:CSA}
Given a text $T$ of length $n$ over alphabet $[1..\sigma]$ and a constant $\epsilon>0$, we can build a data structure which occupies $O(n\log\sigma)$ bits of space and that returns $\mathtt{SA}[i]$ for any $i\in[1..n]$ in time $O(\log^\epsilon n)$. 
\end{lemma}


All the above-mentioned text indexes can trivially be extended to support the same type of queries on a collection of documents instead of a single document. More precisely, given a collection of texts $T_1,T_2,\ldots,T_D$ over the same alphabet $\Sigma$, the same queries can be supported by constructing an index of the string $T_1\$T_\$\ldots T_D\$$. 

\subsection{Colored range reporting and document retrieval}
Muthukrishnan~\cite{muthukrishnan2002efficient} was the first to study the problem of efficiently retrieving documents containing a given string pattern. Through the use of a text index, he reduced the problem to the one of {\em color range reporting}, i.e. reporting all {\em distinct} values (``colors'') occuring in a given interval of an array. His data structure relies on the use of {\em range minimum query} data structures -- a data structure that can find in constant time the smallest element in an sub-range of an array.  
His algorithm was subsequently improved in terms of space (Theorem 4 in \cite{sadakane2007succinct}). 
We will use the following result on color range reporting, which can be obtained by using the optimal range-minimum query data structure~\cite{FH11} in the method of \cite{sadakane2007succinct}:

\begin{lemma}\label{lemma:doc_retrieval}
Given an array $A[1..n]\in [1..\sigma]^n$, we can build a static data structure that occupies $2n+o(n)$ bits that allows reporting all $d$ distinct values occurring in a query interval $A[i..j]$ in time $O(d)$ ($O(1)$ time per reported value). The query will make read-only access to the data structure, read-only random access to elements of the array $A$ 
and read-write access to a bitvector $B$ of size $\sigma$. The bitvector 
needs to be initalized to zero before the first query and is reset to zero at the end of each query. 
\end{lemma}


In combination with text indexing, colored range reporting allows supporting document retrieval queries. More precisely, define the {\em document array} as follows: given a collection of $D$ documents $T_1,T_2\ldots T_D$ of total length $n$, lexicographically sort all the suffixes of the text $T^*=T_1\$T_2\$\ldots T_D\$$, and set $A[i]=j$ iff the suffix of $T^*$ of lexicographic rank $i$ starts inside $T_j$ 
(if the suffix starts with $\$$, then set $A[i]=0$). Document array $A$ can be easily obtained from a text index of $T^*=T_1\$T_2\$\ldots T_D\$$. 
For this, one can construct a bitmap of length $|T^*|$ with $1$'s at positions of $\$$ in $T^*$ and $0$'s otherwise. Then $A[i]=\mathtt{rank}_1(A,\mathtt{SA}[i])+1$ for $i>D$ and $A[i]=0$ for $i\leq D$. 
It is then immediate that using these data structures, Lemmas~\ref{lemma:textindex},~\ref{lemma:CSA}, and \ref{lemma:doc_retrieval} lead to solving the document retrieval problem in time $O(|p|+d\log^\epsilon n)$, where $d$ is the number of resulting documents. For this, we can use the document alphabet-independent BWT index to compute the range $[i..j]$ of occurrences of $p$ in $O(|p|)$ time and then report the $d$ distinct documents that appear in the range $A[i..j]$ in $O(d\log^\epsilon n)$ time.

%% file: Muthukrishnan_based_solution.tex
\section{Solution based on Muthukrishnan's data structure}
\label{muthu-solution}
Our first solution will be 
a combination of tools presented in the previous section. 
We first build a text index for the concatenation of documents $T_1\$T_2\ldots T_D\$$. More specifically, 
we build an instance of the text index of Lemma~\ref{lemma:textindex} which occupies $n\log\sigma(1+o(1))$ bits 
and allows to locate the interval of all suffixes 
of the documents that start with $p$ in time $O(|p|)$. We also build the document array $A[1..n]$, of size $n\log D$, indexed by the document suffixes sorted in lexicographic order and storing the documents each of the suffixes belongs to. 

We further store $h$ instances $C_1,\ldots C_h$ of the data structure of Lemma~\ref{lemma:doc_retrieval},
one instance per level of the tree, defined as follows. 
Consider $d$ (virtual) arrays $A_i[1..n]$, one per level $i\in[1..h]$ of the tree, such that
$A_i[j]$ stores the ancestor at level $i$ of document $A[j]$. Then, each $C_i$ is the data structure of Lemma~\ref{lemma:doc_retrieval} for supporting range reporting queries on array $A_i$. 
Thus, $C_i$ allows to return, for any interval $[r..\ell]$, all distinct elements in $A_i[r..\ell]$ in constant time per element provided that a random-access to each element in $A_i$  is supported in constant time. 

Note that according to Lemma~\ref{lemma:doc_retrieval}, a query will need to use $D$ bits of working space\footnote{We define the working space as a writable space that is only used during queries and is restored to its initial state at the end of the query} 
since it will need to use a temporary bitvector $B$ of size $D_i\leq D$ where $D_i$ is the number of nodes at level $i$ of the tree\footnote{We can use the same bitvector $B$ (Lemma~\ref{lemma:doc_retrieval}) of size $D$ for all $h$ levels: for a query on level $i$, the first $D_i$ bits of $B$ are initally set to zero and are reset to zero at the end of the query}. By Lemma~\ref{lemma:doc_retrieval}, each $C_i$  occupies only $2n+o(n)$.
Finally, in order to simulate constant-time random access to entries of arrays $A_i$, $1\leq i\leq h$, we build a data structure for constant-time level ancestor queries on the category tree (Lemma~\ref{lemma:laq}). Notice that we can access cell $A_i[j]$ using the formula $A_i[j]=\mathtt{LAQ}(\mathtt{leafselect}(A[j]),i)$.
The data structure will occupy $2\Delta+o(\Delta)$ bits of space, where $\Delta$ is the total number of nodes in the tree. 

To answer a query consisting of a pattern $p$ and level $i$, we proceed as follows. We first compute, in time $O(|p|)$, the interval $[\ell..r]$ of suffixes using the BWT-index (Lemma~\ref{lemma:textindex}). The documents containing $p$ are then those contained in $A[\ell..r]$. We then have to output all distinct ancestors at level $i$ of documents $A[\ell..r]$, i.e. all distinct elements of $A_i[\ell..r]$. This is done in constant time per reported element using $C_i$, as follows from Lemma~\ref{lemma:doc_retrieval} and constant-time access to elements of $A_i$ using $\mathtt{LAQ}$ and $\mathtt{leafselect}$ queries. 

The document array occupies $n\log D$ bits of space. The text index is built on top of the $n\log\sigma(1+o(1))$ bits. 
Each of the $h$ instances of the data structure of Lemma~\ref{lemma:doc_retrieval} will occupy $2n+o(n)$ bits of space each for a total space of $2nh+o(hn)$ bits of space. The data structure built on top of the category tree occupies $2\Delta+o(\Delta)$ bits of space. 

We thus have proved the following theorem: 
\begin{theorem}
	\label{muthu-theorem}
Given a collection of $D$ documents of total length $n$ over alphabet $[1..\sigma]$ so that the documents 
are organized in a hierarchy of documents represented by a tree of total size $\Delta$ 
and of height $h$, we can build a data structure 
of size $n(\log\sigma(1+o(1))+\log D+O(h)) + O(\Delta)$ bits of space that, given a pattern $p$, can find all $t$ categories of documents at a given level $i$ that have at least one document that contains the pattern in total time $O(|p|+t)$. 
\end{theorem}
This data structure will be good enough whenever $h$ is small, for example, when $h=\log D$, which holds for example when each internal node in the tree has at least two children.

%% file: Wavelet_tree_based_solutions.tex
\section{Wavelet-tree-based solution}
\label{wavelet-tree-based}
If each node of our tree is branching, i.e. has two or more children, then $h=O(\log D)$ and the solution of Secton~\ref{muthu-solution} takes  $O(n(\log\sigma+\log D))$ bits of space. (Recall that all leaves of our tree occur at level $h$) However, this may not be the case as the tree may have many non-branching (unary) nodes. In the extreme case, we may have $h=\Omega(D)$ and the space of Theorem~\ref{muthu-theorem} will become $\Omega(nD)$ which can be too large if $D$ is large. 
In this section, we deal with this issue and present solutions based on wavelet trees. 

As in Secton~\ref{muthu-solution}, we assume that we first located an interval $[\ell..r]$ in the document array $A$ that corresponds to the occurrences of the query pattern $p$. The goal is then to return 
all internal nodes at level $i$ containing documents from $A[\ell..r]$ in their subtree. 
In Section~\ref{basic-wavelet}, we present the first ''warm-up'' solution that we subsequently improve in Section~\ref{heavy-path-solution}.

\input{Wavelet_tree_based_solution1.tex}

\input{Wavelet_tree_based_solution2.tex}

%% file: Wavelet_tree_based_solution1.tex
\subsection{Basic wavelet-tree-based solution}
\label{basic-wavelet}

We build our wavelet tree on top of the input tree representing the hierarchy of the documents. Therefore, our initial wavelet tree is generally non-binary and non-balanced. As does the input tree, our wavelet tree has height $h$ and $O(\Delta)$ nodes in total. To save space, we will eliminate unary nodes from the wavelet tree (such a node $\alpha$ stores a trivial sequence $C_\alpha=1^{|S_\alpha|}$, see Section~\ref{bitvecs}) and only encode $O(D)$ branching nodes. For each branching node $\alpha$ we store its depth denoted $\delta_\alpha$. Besides the wavelet tree, we will need a data structure for level ancestor queries (Lemma~\ref{lemma:laq}) for the input tree that occupies $O(\Delta)$ bits of space and answers queries in constant time. 

Our alphabet $\Sigma$ will be defined to be the set of documents $[1..D]$. The alphabet interval $\Sigma_\alpha$ assigned to a node $\alpha$ will be the indices of documents occurring in the subtree rooted at $\alpha$. 
The string $S$ for which the tree is built will be the document array $A[1..n]$. 

Our wavelet tree may have nodes with more than two children and we implement them by local {\em binarization}. 
If a node has $d$ children, we will encode it using a binary wavelet tree of $\log d$ levels, called a {\em local wavelet tree}. 
In total, the wavelet tree occupies $n(h\log D)$ bits, since the tree contains $h$ levels and each of the $n$ elements of the document array will contribute at most $\log D$ bits to each level. 

Consider now a query which is defined by a pattern $p$ and a level $i$ in the input tree. Once we computed the document array interval corresponding to $p$, say $A[\ell..r]$, we use our wavelet tree to identify the desired nodes at level $i$. Starting from the root, we traverse the tree top-down through all the nodes $\alpha$ whose assigned sub-alphabet $\Sigma_\alpha\subseteq [1..D]$ intersects with elements of $A[\ell..r]$. This is done by recomputing the current interval for each traversed node. An invariant of this computation is that querying a node $\alpha$ with an interval $[i..j]$ ensures that all elements of $A[\ell..r]\cap \Sigma_\alpha$ are within $S_\alpha[i..j]$. Interval computation is done using $\mathtt{rank}$ queries on binary vectors $B_\alpha$ stored at nodes $\alpha$ of the wavelet tree, we refer to \cite{gagie2009range} where this computation is described in detail. 
We stop the traversal at a node $\alpha$ as soon as $\delta_\alpha\geq i$ and report its ancestor at level $i$ using the level ancestor data structure.

%
The original tree has at most $h$ levels and each node is replaced by a local wavelet tree with at most $\log D$ levels, therefore 
a root-to-leaf path in the wavelet tree has at most $h\log D$ nodes, and the total worst-case query time will be $O(h\log D)$ per 
reported node. 

We now analyse the space usage of the data structure. Since the wavelet tree has $D$ leaves and all nodes are branching, the total number of nodes is $O(D)$. 
Thus, the total space used by the wavelet trees is $n(h\log D)(1+o(1))+O(D\log n)$ bits (see Lemma~\ref{lemma:WT}). The space used by the BWT-index is $n\log\sigma(1+o(1))$ (Lemma~\ref{lemma:textindex}) and the space used by the document array is $n\log D$ bits. The space used by the data structure for level-ancestor queries is $O(\Delta)$ bits (Lemma~\ref{lemma:laq}). 
We thus proved the following theorem.
\begin{theorem}
Given a collection of $D$ documents of total length $n$ over alphabet $[1..\sigma]$ and so that the documents 
are organized in a hierarchy of documents represented by a tree of height $h$, 
we can build a data structure of size $n(\log\sigma+(h+1)\log D)(1+o(1))+O(D\log n)+O(\Delta)$ bits of space that can, given a pattern $p$, find all $t$ categories of documents at level $i$ that have at least one document that contains the pattern in total time $O(|p|+t\cdot h\log D)$. 
\end{theorem}

%% file: Wavelet_tree_based_solution2.tex
\subsection{Solutions based on heavy path decomposition}
\label{heavy-path-solution}
We now describe a more sophisticated solution based on the {\em heavy path decomposition} \cite{sleator1983data,HarelT84} of the wavelet tree from the previous section. Here we present a high-level description of our algorithms, full details will be given in the extended version of the paper.

There are several variants of the definition of heavy path decomposition, with slight differences between the variants. 
In what follows we will use the following variant. With each node $\alpha$ of a given tree $T$, we associate a {\em weight} $w(\alpha)$ equal to the number of leaves in the subtree rooted at $\alpha$. The {\em heavy child} $\beta$ of $\alpha$ is the child of $\alpha$ with the greatest weight, with ties resolved arbitrarily. The other children of $\alpha$ are called {\em light}.
The edge between $\alpha$ and its heavy child is called a {\em heavy edge}, whereas all the other edges from $\alpha$ to its children are called {\em light edges}. 

The heavy path decomposition of a tree $T$ is a decomposition of $T$ into paths defined recursively as follows. We first compute the heavy path (i.e. a path consisting of heavy edges) from the root of $T$ to a leaf, and then recursively apply the decomposition to all subtrees rooted at all light children of the heavy path nodes. 
An interesting property of the heavy path decomposition is that the number of light edges on any root-to-leaf path is at most $\log D$, where $D$ is the number of leaves in the tree. 

\subsubsection{First solution based on heavy path decomposition}
Our first solution will be neither space- nor time-optimal. For each heavy path starting at a node $\alpha$ for which the number of light children of nodes of the path is $\ell_\alpha$, the alphabet will be of size $\ell_\alpha$. We can order the nodes (light children) by increasing depths. The sequence $S_\alpha$ that is associated with a heavy path $\alpha=\alpha_1,\ldots\alpha_k$,
will be of length $n_\alpha$ over alphabet $[1..\ell_\alpha]$, where $n_\alpha$ is the total number of occurrences of leaves (documents) in the subtree rooted at $\alpha$ in the document array $A$. That is, the sequence will be a subsequence of $A[1..n]$, where only the documents that belong to the leaves under $\alpha$ are kept, and the encoding of each element in the subsequence will be the index of the (light) children of the heavy path nodes under which the document appears. Let the depths of the nodes in the heavy path be denoted by $d_1<\ldots<d_k$. We additionally store a bitvector $B_\alpha$ marking the node depths of the different nodes. That is, we initialize the bitvector $B_\alpha$ by all zeros and then set $B_\alpha[d_i]=1$ for every $i\in [1..k]$. 

A query for level $i$ will now proceed as follows. We traverse the tree top-down. For each heavy path, we do the following. 
\begin{enumerate}
\item We first use the bitvector that marks the node depths to determine a subrange $[1..r]$ of the alphabet that will be used for the query (the light nodes included in the range will have depths at most $i$, whereas the nodes in the range $[r+1..h]$ will have depth more than $i$). 
\item We traverse the wavelet tree of the current heavy path. Such a query will spend time $O(t\log\ell_\alpha)$ for a heavy path with $\ell_\alpha$ light children, in which $t$ distinct light children appear in the sequence. 
\end{enumerate} 

It is easy to see that the total space will be $O(n\log^2 D)$ bits, since the alphabet size is $O(\log\ell_\alpha)$ for each node $\alpha$ with $n_\alpha$ stored elements and each element of $A$ will incur at most $\log D$ elements in the wavelet trees stored in the heavy paths of the tree. The query time can be bounded to be $O(\log^2 D)$ per reported document by a similar argument (we traverse $\log D$ heavy paths and each traversal costs $\log D$ time). 

We thus obtain the following result.
\begin{theorem}
Given a collection of $D$ documents of total length $n$ over alphabet $[1..\sigma]$ so that the documents 
are organized in a hierarchy of documents represented by a tree of total size $\Delta$, 
we can build a data structure 
of size $O(n\log^2 D+\Delta)$ bits of space that can, given a pattern $p$, find all $t$ categories of documents at level $i$ that have at least one document containing the pattern in total time $O(|p|+t\log^2 D)$. 
\end{theorem}

\subsubsection{Second solution based on heavy path decomposition}
Our second solution based on heavy path decomposition will rely on a more fine-grained encoding. We will make use of Huffman-shaped wavelet tree~\cite{foschini2006indexing} for each heavy path, such that the wavelet tree node corresponding to a light node of relative weight $w$ (the weight light node divided by weight of the root of heavy path)  will be encoded using $\log(1/w)+O(1)$ bits and the correponding wavelet tree leaf will be at depth $\log(1/w)+O(1)$. It is now easy to see that the encoding of each element of $A$ will take $O(\log D)$ bits and, furthermore, the cost of a query can be upper-bounded by  just $O(\log D)$. Both bounds rely on a telescoping argument. We have the following result. 
\begin{theorem}
Given a collection of $D$ documents of total length $n$ over alphabet $[1..\sigma]$ and so that the documents 
are organized in a hierarchy of documents represented by a tree of total size $\Delta$, we can build a data structure 
of size $n(\log\sigma(1+o(1))+O(\log D))+O(\Delta)+O(D\log n)$ bits of space that can, given a pattern $p$, find all $t$ categories of documents at a level $i$ that have at least one document that contains the pattern in total time $O(|p|+t\log D)$. 
\end{theorem}

%% file: succinct_compressed.tex
\section{Compact and compressed data structures for categorical data queries}
\label{succinct}
In this section we explore more space-effcient versions of the problem. More in detail, we are interested 
in studying the problem under succinct and compressed-space constraints. Namely, our aim is to use 
$O(n\log\sigma)$ bits for the succinct case and $nH_0+o(n\log\sigma)+O(D\log n)$ bits of space for the compressed case. 
To achieve this, we will improve the solution of Section~\ref{muthu-solution}. More precisely, we avoid the storage of the document array and simulate direct access 
to the document array using Lemma~\ref{lemma:CSA}. As a consequence, we can achieve time $O(\log^\epsilon n)$ 
to get the 
given document index $A[i]$ for any $i\in[1..n]$. This will reduce the space to represent the document array from $O(n\log D)$ to $O(n\log\sigma)$ bits. Now the space used by the range minimum query data structures will become the bottleneck. To reduce the space usage we will make use of sparsification. More precisely, we will divide the document array into blocks and sample just the values of the $A$ array that are the smallest in each block. The space becomes $O(n/\alpha)$ bits where $\alpha$ is the sparsification factor. For details on how the sparsification is used to simulate the reporting of distinct documents that appear in interval $A[i..j]$, we refer the reader to~\cite{belazzougui2013improved,hon2014space}. Here we just mention that the time per reported document becomes $O(\alpha\log^\epsilon n)$ and entails $O(\alpha)$ accesses to the document array, each of which requires $O(\log^\epsilon n)$ time. 
We thus have the following result.
\begin{theorem}
Given a parameter $\alpha\geq 1$ and a collection of $D$ documents of total length $n$ over alphabet $[1..\sigma]$ and so that the documents 
are organized in a hierarchy of documents represented by a tree of height $h$, we can build a data structure 
of size $O(n\log\sigma)+O(nh/\alpha)$ bits of space that can, given a pattern $p$, find all $t$ categories of documents at level $i$ that have at least one document that contains the pattern in total time $O(|p|+t\cdot\alpha\log^\epsilon n)$. 
\end{theorem}

By setting $\alpha=\lceil\frac{h}{\log\sigma}\rceil$ we get space $O(n\log\sigma)$ bits and query time $O(|p|+(t+1)\log^\epsilon n\cdot(1+\frac{h}{\log\sigma}))$. 
We thus have the following corollary. 

\begin{corollary}
Given a parameter $\alpha$ and collection of $D$ documents of total length $n$ over alphabet $[1..\sigma]$ and so that the documents 
are organized in a hierarchy of documents represented by a tree of height $h$, we can build a data structure 
of size $O(n\log\sigma)$ bits of space that can, given a pattern $p$, find all $t$ categories of documents at level $i$ that have at least one document that contains the pattern in total time $O(|p|+(t+1)\cdot\log^\epsilon n(1+\frac{h}{\log\sigma}))$. 
\end{corollary}

Whenever $h=\log D$ (e.g. every internal node is branching), the query time simplifies to $O(|p|+(t+1)\cdot\log_\sigma D\cdot\log^\epsilon n)\in O(|p|+(t+1)\log^{1+\epsilon}n)$. 
We can also get compressed space. Namely, we can use a compressed suffix array~\cite{grossi2003high} with query time $\log n\log\log n$ and space $nH_k+o(n)$ to represent the document array. We will combine the compressed suffix array with the alphabet-independent variant of BWT-index presented in~\cite{belazzougui2014alphabet}. We then get an index that uses space $nH_k+o(n\log\sigma)$ with query time $O(|p|)$ to find the suffix array interval of a pattern and $O(\log n\log\log n)$ time to access an element of the suffix array. Notice that we can translate access to a suffix array element to an access to a document array element using $O(D\log n)$ bits of space. Summing up, we get the following theorem. 
\begin{theorem}
Given a parameter $\alpha$ and a collection of $D$ documents of total length $n$ over alphabet $[1..\sigma]$ and so that the documents 
are organized in a hierarchy of documents represented by a tree of height $h$, we can build a data structure 
of size $nH_k+o(n\log\sigma)+O(D\log n)+O(nh/\alpha)$ bits of space that can, given a pattern $p$, find all $t$ categories of documents at level $i$ that have at least one document that contains the pattern in total time $O(|p|+t\cdot\alpha\log n\log\log n)$. 
\end{theorem}
By setting $\alpha=h\cdot \log\log n$, we get space $nH_k+o(n\log\sigma)+O(D\log n)$ bits and query time $O(|p|+t\cdot h\log n(\log\log n)^2)$. The latter becomes $O(|p|+t\log D\log n(\log\log n)^2)$ whenever $h=O(\log D)$.

%% file: conclusions.tex
\section{Conclusions}
In this paper, we proposed several solutions for the problem of categorical retrieval. Possible extensions of our work include the case when the document hierarchy is a DAG rather than a tree. This situation occurs, for example, with phylogenetic networks. 
The solution in Section~\ref{muthu-solution} could easily be extended to DAG structured categories if there was an efficient support for level ancestor queries on DAGs. Other possible extensions includes top-k queries in which categories are either ordered by a static order or by the total frequency of the pattern in the documents that belong to the reported categories.